\documentclass{article}
\usepackage{hiph-art}
\usepackage{epsfig}
\newcommand{\be}{\begin{eqnarray}}
\newcommand{\ee}{\end{eqnarray}}
\volnumber{19} \issuenumber{1} \edyear{2004}                             
\frompage{000} \topage{000}                                              
\recrevdate{1 January 2004}                                              
\def\rightmark{  From chemical freezeout to critical conditions in heavy ion collisions}
\def\leftmark{Krzysztof Redlich}

\title{    From chemical freezeout to critical conditions in heavy ion collisions}
\authors{
{Krzysztof Redlich$^{1,2}$ %
\index{One, A.} 
\index{Two, A.} 
}\\[2.812mm]
{\normalsize
\hspace*{-8pt}$^1$ Institute of Theoretical Physics University of Wroclaw,\\
PL--50204 Wroc\l aw, Poland\\
\hspace*{-8pt}$^2$ Fakult\"at f\"ur Physik, Universit\"at Bielefeld,\\
\small Postfach 100 131, D-33501 Bielefeld, Germany\\
[0.2ex]
}}

\abstract{ We compare the statistical thermodynamics of hadron
resonance gas with recent LGT results
 at finite chemical
 potential.  We argue that for $T\leq T_c$ the equation of state derived
 from
 Monte--Carlo simulations of two quark--flavor QCD at finite
chemical potential is consistent with that of
 a hadron resonance gas when applying  the same set of
approximations as used in LGT calculations.
 We indicate  the
relation of chemical freezeout conditions obtained from a detailed
analysis of particle production in heavy ion collisions with the
critical conditions required for deconfinement. We argue that the
position of a hadron--quark gluon boundary   line in temperature
chemical potential plane  can  be  determined in terms of the
resonance gas model by the condition of fixed energy density.}

\PACS{ }

\makeindex
\begin{document}

\maketitle
\section{Introduction}
A detailed analysis of particle production in heavy ion collisions
has shown that in a broad energy range from SIS through AGS,
SPS\footnote{ The most recent thermal analysis of SPS data was
presented in Ref. \cite{bec}.} up to RHIC particle yields resemble
that of chemical equilibrium population with respect to the
partition function of hadron resonance gas (HRG) \cite{rev}.

\begin{figure}[htb]
 {\hskip -.3cm
\includegraphics[width=28.8pc]{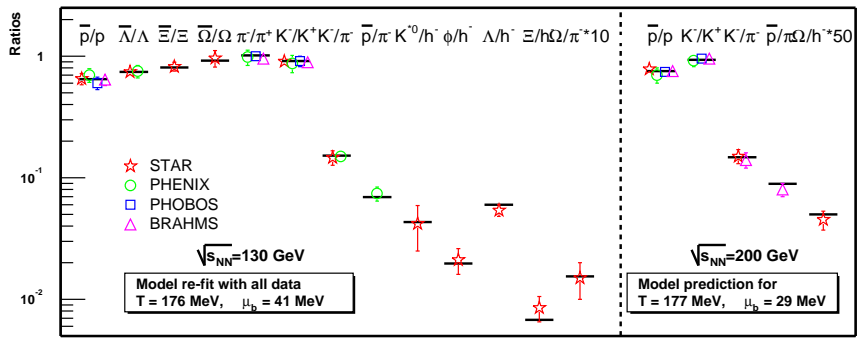}}\\
\vskip -.5cm { \caption{ \label{ff3} Comparison of the
experimental data on different particle multiplicity ratios
obtained  at RHIC at $\sqrt s_{NN} =130$  and 200 GeV  with
thermal model calculations. The thermal model analysis is from
Refs.~\protect\cite{r10,r10n}.
 }}
\end{figure}

 In Fig. (\ref{ff3}) we show an example of the comparison
 of HRG--model with experimental data on different  particle yields ratios
 measured in Au--Au collisions at RHIC
 energies. The model results \cite{r10,r10n} are shown as horizontal lines whereas
 experimental data are indicated as points. The apparent quite
 good agreement of the model predictions with the experiment indicates
 that  production of particles in heavy ion collisions is of thermal origin \cite{larry}.
 A collision  fireball at chemical freezeout is characterized by only
 two thermal parameters, the temperature $T$ and baryon chemical
 potential\footnote{ The quark chemical potential $\mu_q={1\over
3}\mu_B$} $\mu_B$. The volume of the fireball can be possibly
connected with the HBT source size \cite{ceres,satz}.

\begin{figure}[htb]
 {\vskip -0.5cm
{\hskip 0.6 cm
\includegraphics[width=22.5pc, height=21.5pc,angle=-180]{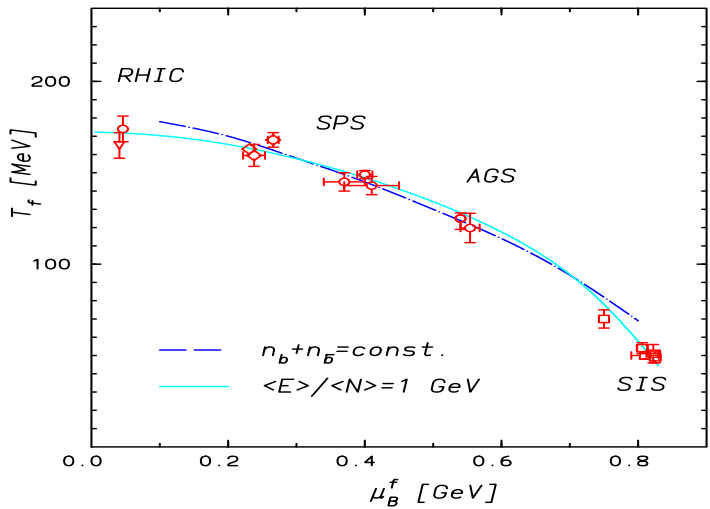}}}\\
{\vskip -1.cm \caption{\label{ff200} Chemical freezeout in heavy
ion collisions.  The full line represents the condition of  fixed
energy/particle $\simeq 1.0$ GeV from Ref. \protect\cite{r9}.  The
broken line describes the chemical freeze--out conditions of
fixed total density of baryons
 plus antibaryons, $n_b+n_{\bar b}=0.12/$fm$^3$ from
 Ref.~\protect\cite{rpb}. The points are the average values of
 $T_f$ and $\mu_B^f$ obtained from the analysis of particle yields
 in heavy ion collisions at the SIS, AGS, SPS and RHIC energy
 \protect\cite{rev}.
  }}\label{ff20}
\end{figure}

  Figure~(\ref{ff20}) shows the  compilation of the chemical freeze--out
parameters that are required to reproduce the measured particle
yields in central Au-Au or Pb--Pb  collisions at SIS, AGS, SPS and
RHIC energy. The GSI/SIS results have the lowest freeze--out
temperature and the highest baryon chemical potential. As the beam
energy increases a clear shift towards higher $T$ and lower
$\mu_B$ occurs. There is a common feature to all these points,
namely that the average energy $\langle E\rangle$ per average
number of hadrons  $\langle N\rangle$   is approximately 1 GeV. {
A chemical freeze--out} in A--A collisions is thus reached
\cite{r9} { when the energy per particle $\langle
E\rangle$/$\langle N\rangle$ drops below 1 GeV} at all collision
energies. The physical origin of  the above freezeout condition
requires  dynamical justification. Recently, this point has been
investigated in central  Pb--Pb collisions at the SPS  in terms of
the Ultra--relativistic Quantum Molecular Dynamics model
(UrQMD) \cite{r72,r72n}. 
 A detailed
study has shown that there is a clear correlation between the
chemical break--up in terms of inelastic scattering rates and the
rapid decrease in energy per particle. If $\langle
E\rangle/\langle N\rangle$ approaches the value of  1 GeV the
inelastic scattering rates drop substantially and  further
evolution is due to elastic and pseudo-elastic collisions that
preserved the chemical composition of the collision fireball.
Following  these  UrQMD results  one could consider the
phenomenological chemical freeze--out of $\langle E\rangle/\langle
N\rangle\simeq 1$ GeV as the condition of inelasticity in heavy
ion collisions.

 Chemical freeze--out  in heavy ion collisions was recently
proposed to be determined  \cite{rpb} by the condition of  fixed
density of the total number of baryons plus  antibaryons $F\equiv
(n_B+n_{\bar B})\simeq 0.12/fm^3$. It is seen in
Fig.~(\ref{ff200}) that, within statistical uncertainties on the
freeze--out parameters the above condition provides a good
description of experimental data from the top AGS up to RHIC
energy. However, at the low  SIS energy it overestimates the
freeze--out temperature for a given chemical potential.
Consequently, e.g. the yield of pion/proton or pion/participant
and strange/non--strange particle ratios obtained at SIS turn out
to be larger than the experimental values. Taking the value of
$F\sim 0.05/fm^3$ at SIS energy and using the freezeout condition
of $F_{SIS}=fixed$ leads eg. at RHIC to the freezeout temperature
of $\sim$150 MeV, the value which is lower than obtained from the
particle yield ratios as seen in Fig. (\ref{ff20}).  At $\mu_B=0$
the $\langle E\rangle/\langle N\rangle$ condition converge to
$T_f\simeq 170$ MeV which is well consistent with the critical
temperature $T=167(13)$ \cite{tn} and $T=173(8)$ \cite{pai}
expected in (2+1) and 2--flavour QCD respectively.

 At SPS and RHIC the freeze-out
parameters, the temperature $T_f$ and the energy density
$\epsilon_f$, predicted by the presently used thermal model of
hadron resonance gas, agree well with the values of the
corresponding parameters required for deconfinement in  LGT
calculations \cite{rev,stach,our}. The above quantitative
agreement of freeze-out and critical parameters  suggests that at
SPS and RHIC the chemical freeze-out appears in the near vicinity
of the phase boundary \cite{stf}. If this is indeed the case then
the phenomenological statistical operator of hadron resonance gas
$Z_{HG} $ should also provide, consistent with LGT,  description
of QCD thermodynamics in
 confined, hadronic phase  of QCD \cite{our,maria1,maria2,our1}. In the following  we argue that the basic
qualitative properties of $Z_{HG} $ resulting from its dependence
on $T$ and baryon chemical potential $\mu_B$ are indeed  present
in the recent lattice results of two flavor QCD at the finite
chemical potential.
We also show that  imposing a fixed energy density condition for
deconfinement the hadron resonance gas partition function $Z_{HG}$
can also describe \cite{our} the quark mass and number of flavor
dependence of the lattice critical temperature at $\mu_B=0$ and
the position of the phase boundary   in the $(T,\mu_B)$--plane at
small $\mu_B$ \cite{our2}.
\begin{figure}[htb]
\begin{minipage}[t]{58mm} \hskip .3cm  \vskip -5.16cm

\includegraphics[width=14.1pc,height=12.pc,angle=0]{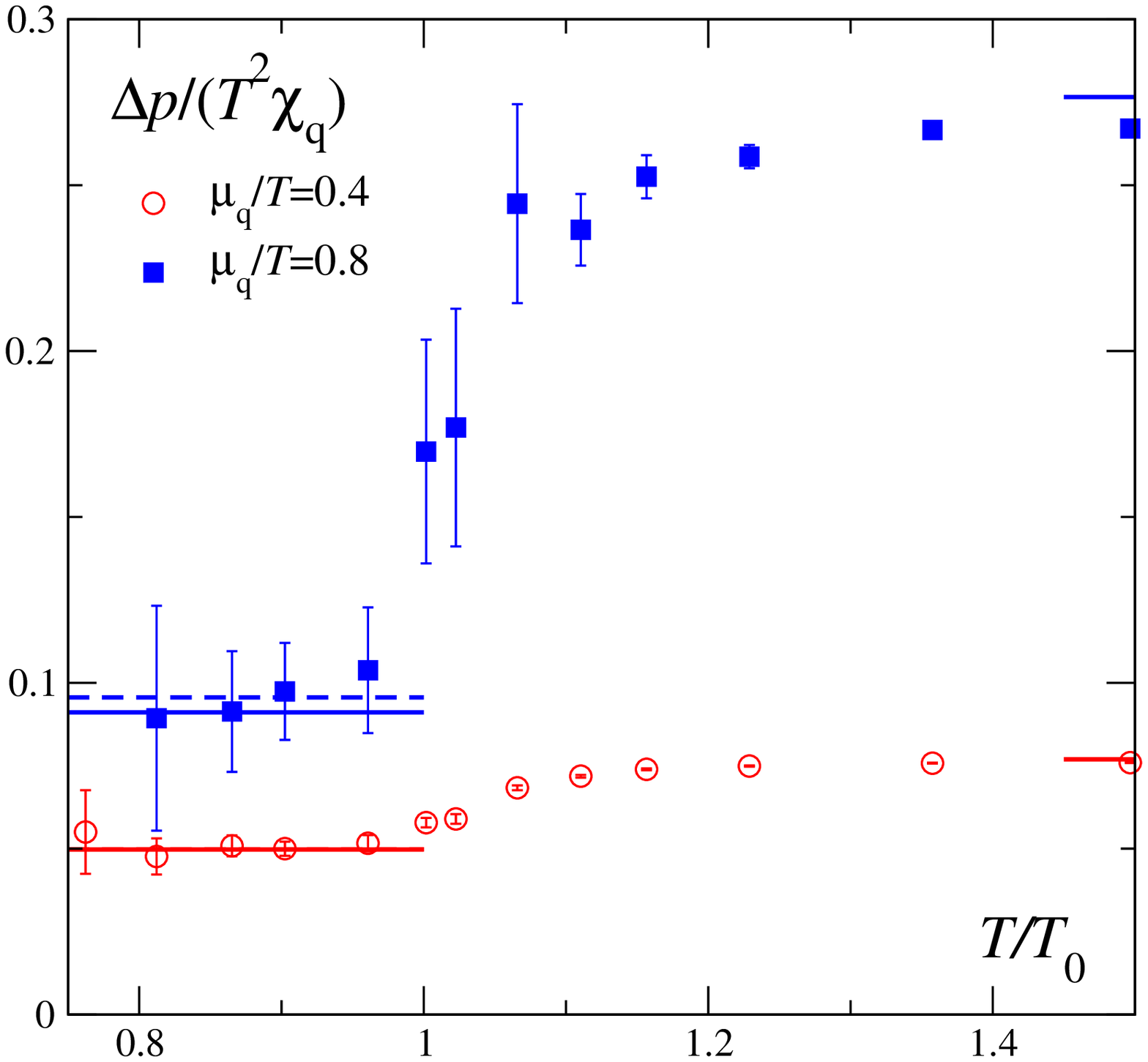}
\end{minipage}
 \hskip 0.5cm
 \begin{minipage}[t]{58mm}
{
\includegraphics[width=14.1pc, height=12.pc,angle=0]{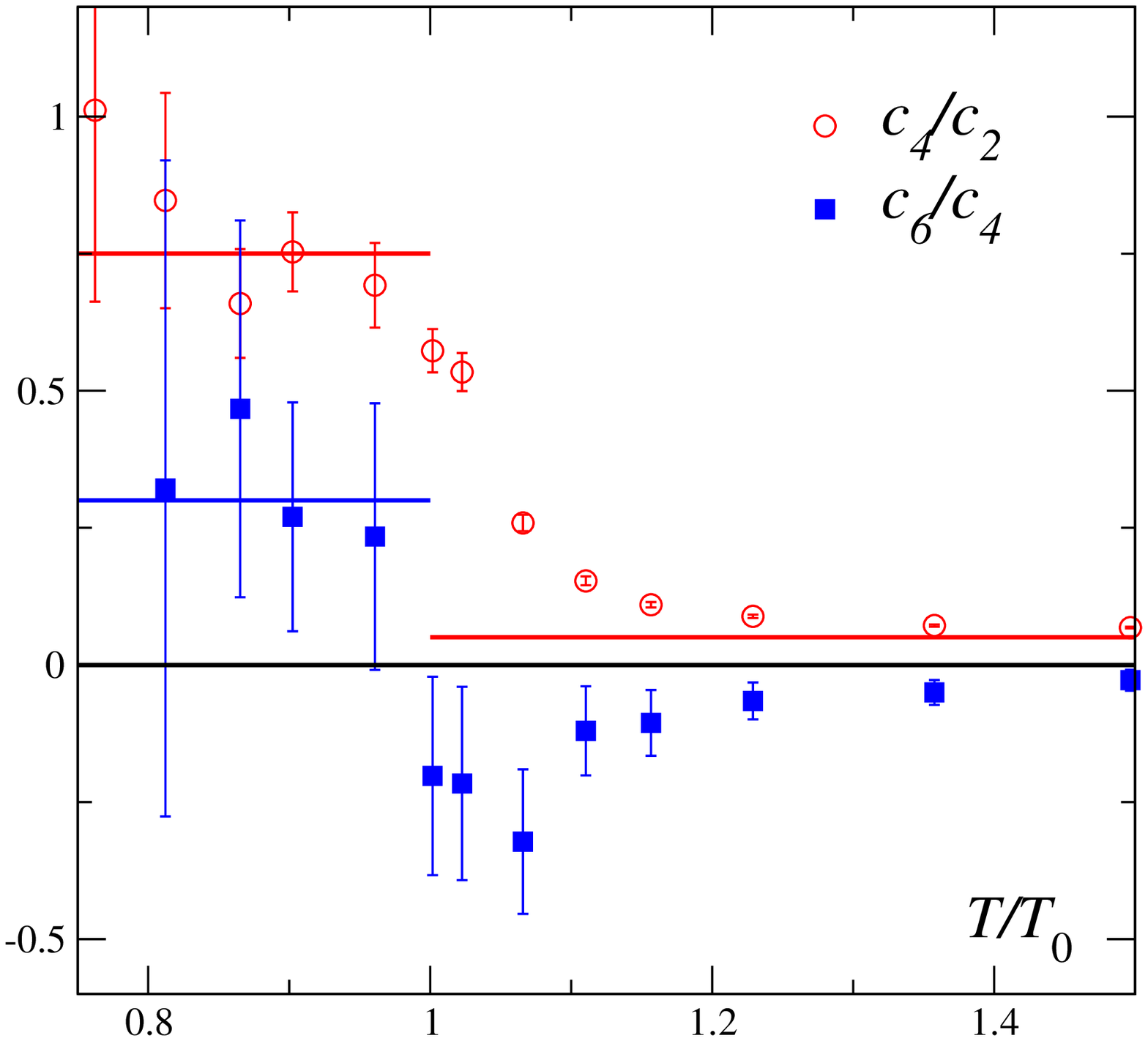}}\\
 \end{minipage} \vskip -0.5cm
\caption{ \label{fig:c2c4} Left--hand figure: temperature
dependence of $\Delta P/(\chi_qT^2)$ ratio  for two different
values of $\mu_q/T$. The 2--flavor  lattice results
 are from  \protect{\cite{lgt,lgt1}}. The lines are
the  hadron resonance gas model values  with (broken--line) and
without (full--line) the Taylor expansion of $\cosh(\mu_B/T))$
\protect{\cite{our}}.   The right--hand side shows the temperature
dependence of the ratios of the second, fourth and sixth order
coefficients in the Taylor expansion of thermodynamic pressure.
The points are lattice results from \protect{\cite{lgt,lgt1}}. The
lines at $T<T_0$ are the values obtained from the expansion of
$\cosh(\mu_B/T))$.
  }
\end{figure}
\section{QCD thermodynamics of hadronic phase}

The phenomenological partition function used in the description of
particle productions in heavy ion collisions was, following
Hagedorn,  constructed as a non--interacting hadronic gas which is
composed of all hadrons and resonances.  In this approach the
interactions between hadrons are included to the extent that the
thermodynamics of an interacting system of elementary hadrons is
effectively approximated by that of mixture of ideal gases of
stable particles and resonances \cite{hagedorn,model}. In the
Boltzmann approximation, suitable for the moderate values of
$\mu_B<m_N$ and $T\geq 50$ MeV, there is a factorization of $T$
and $\mu_B/T$ dependence in  relevant observables characterizing a
baryonic sector of the system. The basic quantity is the pressure
$\Delta P =P(T,\mu_B)-P(T,\mu_B=0)$

\begin{equation}
\hspace*{-1.3cm} {{\Delta P}\over {T^4}}\simeq F(T)(\cosh
({{\mu_B}\over T}-1)~~~~~~{\rm with}~~~~~~ F(T)\simeq \int dm\rho
(m) ({m\over T})^2 K_2({m\over T})\label{eq1}
\end{equation}
from which the baryon density $n_B$ and the baryon number
susceptibility $\Delta\chi_B$ are obtained as the first and second
order derivatives with respect to $\mu_B$, respectively. In the
phenomenological application of the above model the mass spectrum
$\rho(m)$ in Eq. (\ref{eq1}) is approximated as the sum of delta
functions. The baryonic pressure gets the contributions from all
known baryons and resonances.

The net baryonic pressure $\Delta P(T,\mu_q)$ has been recently
obtained on the lattice  in two flavor QCD as the Taylor series in
$\mu_q/T$

\begin{equation}
{\Delta p(T,\mu_q) \over T^4} \simeq \sum_{n=1}^{n=3} c_{2n}(T)
\left( {\mu_q \over T} \right)^{2n} \quad . \label{eq9}
\end{equation}
 up to $O(\mu_q^6)$ order \cite{lgt,lgt1}.

The basic observables characterizing baryonic contribution to the QCD
thermodynamics,  the  net quark  density $n_q$ and susceptibility
$\chi_q$  are obtained from Eq.(\ref{eq9}) as
\begin{eqnarray}
{n_q \over T^3} &=& {\partial\; \Delta P(T,\mu_q)/T^4 \over
\partial\; \mu_q/T} \simeq 2\; c_2(T) \left( {\mu_q \over T} \right)
+ 4\; c_4(T) \left( {\mu_q \over T} \right)^3 \quad  +
 6\; c_6(T) \left( {\mu_q \over T} \right)^5 \quad  \nonumber \\
{\chi_q \over T^2} &=& {\partial^2\; \Delta p(T,\mu_q)/T^4 \over
\partial\; (\mu_q/T)^2} \simeq 2\; c_2(T) + 12\; c_4(T) \left( {\mu_q
\over T} \right)^2 + 30\; c_6(T) \left( {\mu_q \over T} \right)^4
. \quad \quad  \label{eq10}
\end{eqnarray}

The coefficients $c_i(T)$ in the Taylor expansion
(\ref{eq9},\ref{eq10}) were  calculated through the Monte--Carlo
simulations of 2--flavor QCD \cite{lgt,lgt1}.

The lattice results (\ref{eq9}-\ref{eq10}) restricted to the
confined phase of QCD can be directly compared with the
predictions of the hadron resonance gas model (\ref{eq1}). This,
in general, required a numerical analysis. However, there are
particular features of the resonance gas partition function that
could be directly verified by the LGT results.

 First, as
seen from Eq. (\ref{eq1})  there is a factorization of $T$ and
($\mu_q/T$) dependence in the thermodynamic pressure and
associated baryonic observables $n_q$ and $\chi_q$. Thus, in the
resonance gas model  any ratios of these observables, calculated
at   fixed quark chemical potential $(\mu_q/T)=fixed$, should be
independent of temperature \cite{our,our1}. In Fig. (3--left) we
show as an example the ratio of $\Delta P/\Delta (\chi_BT^2)$ for
two different values of $\mu_q/T$  as the function of $T$. It is
clear from Fig. (3--left) that the factorization predicted by HRG
partition function $Z_{HG}$ is also seen in LGT results for
$T<T_c$.

Second, the $(\mu_q/T)$--dependence in the HRG--model is described
by the (cosh)--function. Thus, performing  the Taylor expansion of
the  pressure (\ref{eq1}) in $(\mu_q/T)$

\be
 {{\Delta P(T,\mu_q)}\over {T^4}}   \simeq \; F(T) \left(
{{9}\over 2}
 \;  \left( {\mu_q
\over T} \right)^2 + {{81}\over {4!}} \;  \left( {\mu_q \over T}
\right)^4 + {{3^6}\over {6!}} \;  \left( {\mu_q \over T} \right)^6
\right), \quad \quad
 \label{eq11}
 \ee
one gets the predictions on the values of the coefficients $c_i$
in (\ref{eq10})   from the HRG--model

\be c_2(T)= {{9}\over 2}F(T)\;,\; c_4={{81}\over {4!}}F(T) \;,\;
c_6=
 {{3^6}\over {6!}}F(T). \quad \quad
 \label{eq11a}
 \ee
where for a discrete hadronic mass spectrum

\be
 F(T)=\sum_i
{{d_i}\over {\pi^2}}({{m_i}\over T})^2K_2({{m_i}\over T})
 .\label{eq11b}
 \ee
 with the sum taken over all baryons and baryonic resonances of mass $m_i$
 and the spin--isospin degeneracy factor $d_i$.

In  Fig. (3--right) we show the ratios  $c_4/c_2$ and $c_6/c_4$
obtained on the lattice in 2--flavour QCD and the corresponding
results of the HRG--model from Eq. (\ref{eq11a}): $c_4/c_2=3/4$
and $c_6/c_4=0.3$. The temperature dependence of the Taylor
coefficients  in Eq. (\ref{eq11}) and (\ref{eq11a}) is controlled
by a  common function for all $c_i$. Thus, in   HRG--model, the
ratios of different $c_i$ are independent of $T$. The lattice
results for $T<T_c$ are, within statistical errors, consistent
with this prediction. The values of  $c_6/c_4$  and $c_4/c_2$
ratios obtained recently on the lattice \cite{lgt,lgt1}   are also
seen in Fig. (3--right) to coincide  with  HRG--model. At
temperature $T\simeq T_c$, however, the lattice results deviate
from the resonance gas values. This is to be expected as at the
critical temperature, deconfinement releases the color degrees of
freedom\footnote{These color  degrees of freedom   could be also
possibly described as  resonance bound states \cite{shuryak}.}
which are obviously not present in the statistical operator of the
hadron resonance gas. The partition function (\ref{eq1}) is not
sensitive to the phase transition thus, could be at most
applicable  below  deconfinement.


The pressure calculated on the lattice  (see Fig. (4) ) increases
abruptly when approaching  deconfinement transition from the
hadronic side. If the phenomenological statistical operator
$Z_{HG}$ is of physical significance  then this increase could be
due to the resonance formation. To check the importance of resonances
near deconfinement one would need to reproduce lattice results on
the $T$--dependence of $\Delta P$ at fixed $\mu_q/T$. However, to
quantify this dependence one needs  to implement  the same set of
approximations in Eq.(1) as those being used on the lattice. The
lattice results were obtained in 2--flavor QCD, thus there is no
contribution of strange baryons in Eq.(1). In addition due to the
Taylor expansion of $\Delta P$ in  the lattice calculations one
also needs to apply  a similar approximation in Eq.(1).

 The current lattice results in Fig. (4) were  obtained
 with
 a large quark mass corresponding to $m_\pi\simeq 770$MeV. This
  large quark mass
 distorts the baryon mass spectrum in  Eq. (\ref{eq1}).
 The  pion mass dependence of baryon masses    can
be obtained from lattice calculations at zero temperature
\cite{our,par}. We use the following ansatz for the
parametrization  of baryon mass  dependence on the pion mass,
\begin{equation}
{{m^*(m_\pi)}\over m}\simeq 1+A{{m_\pi }\over {m^2}}, \label{par}
\end{equation}
with $A=0.9\pm 0.1$,  $m^*$ being  a distorted hadron mass at
fixed $m_\pi$ and $m$ is its corresponding physical value.

  In
Fig. (4) the lattice results are compared with Eq.(\ref{eq1})
calculated with modified baryon mass spectrum following Eq.
(\ref{par}). The $T$ dependence of QCD thermodynamics obtained on
the lattice at different values of $\mu_q/T$ is seen in Fig. (4)
to be well consistent with the predictions of a HRG--model. A
similar agreement  was  also found for  other relevant
thermodynamic observables in (2+1), 3--flavour  \cite{our} and
4--flavour QCD \cite{maria1}.
The above agreements indicate   that the phenomenological
partition function of hadron resonance gas should be  a good
approximation of the QCD statistical operator in the hadronic
phase.
 \begin{figure}
\begin{minipage}[t]{58mm}
\hskip -1.3cm  \vskip -2.8cm \hskip -1.cm
\includegraphics[width=35.5pc,height=28.pc,angle=180]{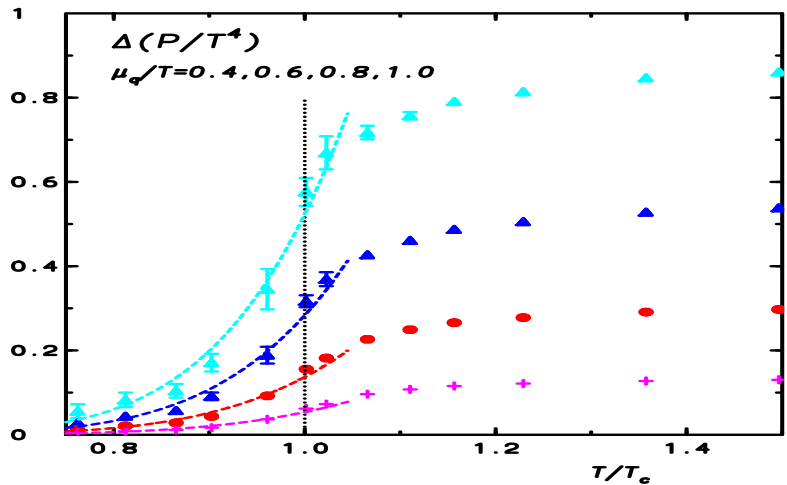}
 \end{minipage}
 \vskip -3.0cm \caption{
 Temperature dependence of baryonic
 pressure for different values of
$\mu_q/T$. The lattice results are from \protect{\cite{lgt}}. The
lines are the statistical model results \protect{\cite{our}}.
 }\label{fig:c2c}
 \end{figure}

\section{Critical conditions for deconfinement}

The recent lattice results on QCD thermodynamics show  that
deconfinement occurs at similar values of the energy density in
QCD with light quarks as well as in the pure gauge theory,
although the transition temperature shifts by about 40\% and
$\epsilon /T_c^4$ differs by an order of magnitude \cite{pai}.
Thus, it has been suggested that for arbitrary quark masses the
transition in QCD appears at the roughly constant energy density. Such
an assumption can be verified by the HRG--model formulated for
an arbitrary value of the quark mass \cite{our}. In Fig.~(5) we show
the lines of the constant energy density calculated in the resonance
gas model and compare these to the transition temperatures
obtained in lattice calculations. As can be seen the agreement is
quite good up to masses, $m_{\pi}\simeq 3\; \sqrt{\sigma}$ or
$m_{\pi}\simeq 1.2$~GeV.  At larger values of the quark mass the
observed deviations  are due to the fact that thermodynamics is
dominated by glueballs which are not present in the hadronic mass
spectrum \cite{our}.

 \begin{figure}
\begin{minipage}[t]{58mm}
\hskip -1.3cm  
\hskip 1.cm
\includegraphics[width=15.pc,height=16.7pc]{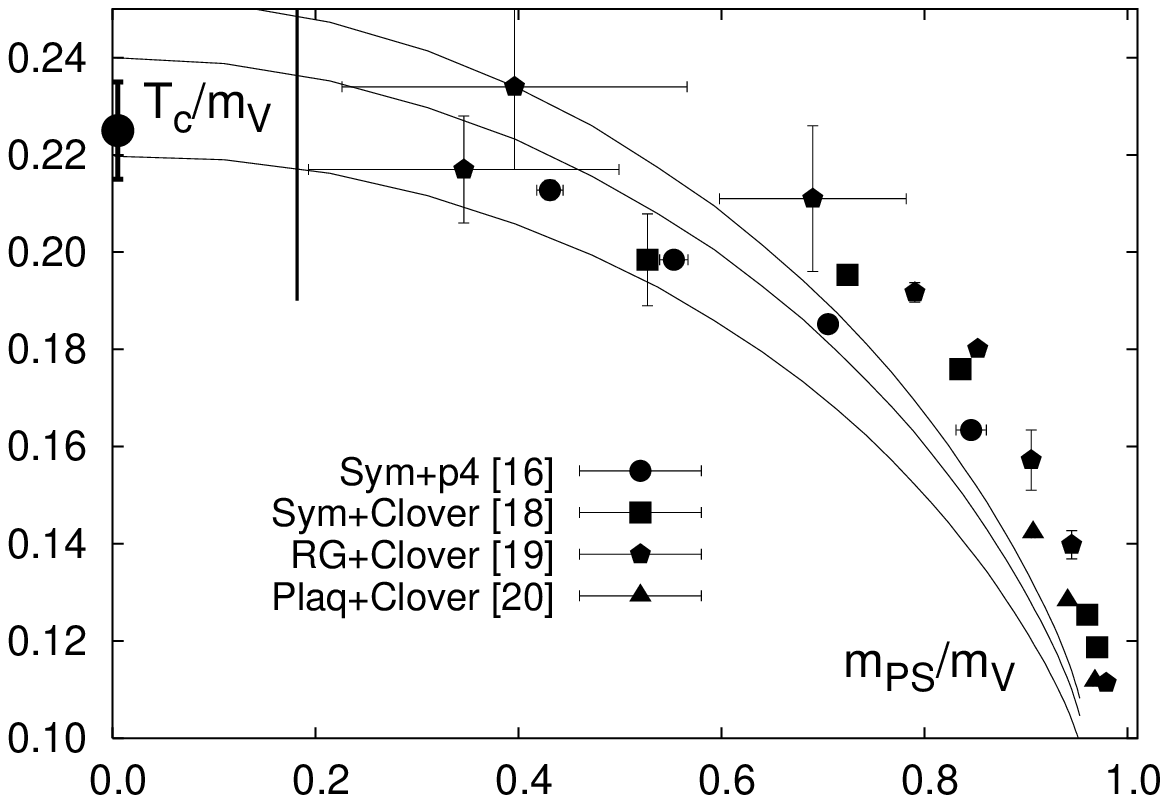}
 \end{minipage}
 \hskip 0.4cm
 \begin{minipage}[t]{58mm}
\includegraphics[width=15.pc,height=16.7pc]{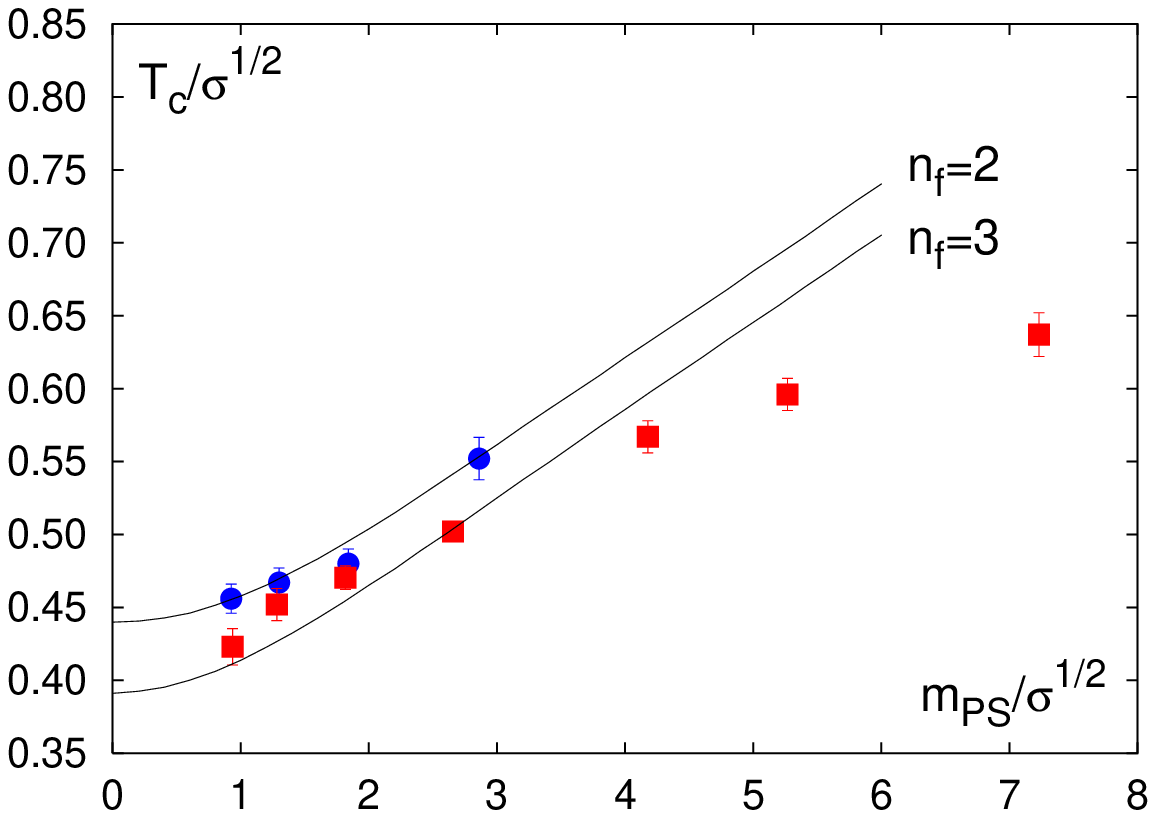}
 \end{minipage}
{\caption {The  transition temperature vs. pion mass obtained in
lattice calculations and lines of constant energy density
calculated in a resonance gas model \protect\cite{our}. The left
hand figure shows a comparison of constant energy density lines at
1.2 (upper), 0.8(middle) and 0.4(lower) GeV/fm$^3$ with lattice
results for 2-flavor QCD obtained with improved staggered
\protect\cite{pai} as well as improved Wilson
\protect\cite{Bernard,Ali,Edwards} fermion formulations. $T_c$ as
well as $m_{PS}$ are expressed in terms of the corresponding
vector meson mass. The right hand figure shows results for 2 and 3
flavor QCD compared to lines of constant energy density of 0.6
GeV/fm$^3$. Here $T_c$ and $m_{PS}$ are expressed in units of
$\sqrt{\sigma}$. For a detailed description see Ref.
\protect\cite{our}.}}
    \label{fig:conditions}
  \end{figure}

 \begin{figure}
 \begin{minipage}[t]{58mm}
\vskip -1.4cm \hskip 0.3cm
\begin{center}
\hskip 0.8cm
\includegraphics[width=28.5pc,height=24.7pc,angle=0]{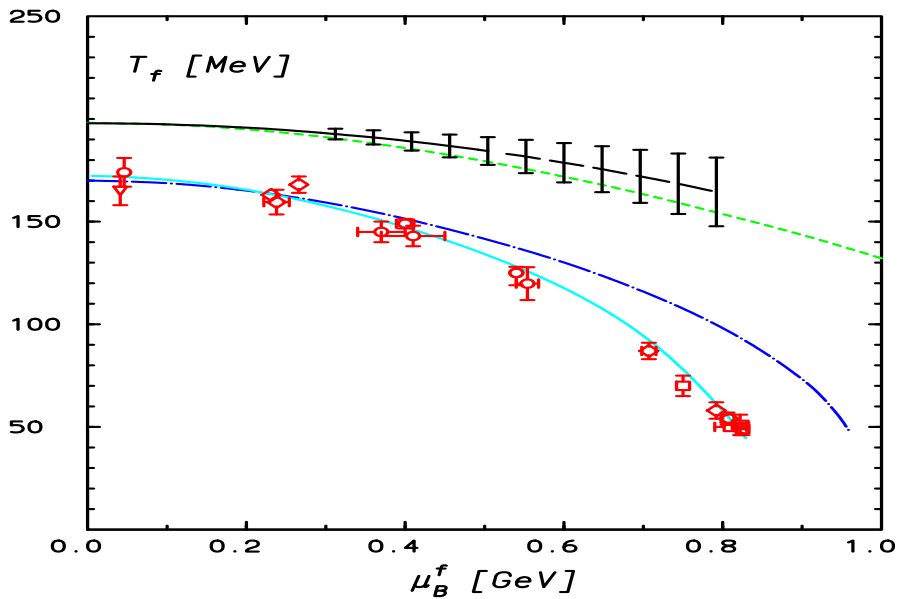}\\
\end{center}
 \end{minipage}
\vskip -1.5cm \caption{ Lattice results \protect\cite{lgt} on the
phase boundary curve (line with errors) together with
phenomenological freeze-out values of $T$ and $\mu_B$ (points)
obtained from the analysis of particle production in heavy ion
collisions \protect{\cite{rev}}. Short-dashed and dashed--dotted
lines are the statistical model results obtained under the
condition of fixed $\epsilon \simeq 0.6$ GeV/$fm^3$ with
$m_\pi\simeq 0.77$ GeV and $m_\pi\simeq 0.14$ GeV respectively.
Also shown (full--line) is the phenomenological freeze-out curve
of fixed energy/particle$\simeq 1$GeV from \protect\cite{r9}.
 }\label{fig:c2c4}
 \end{figure}


The critical conditions of  fixed energy density could be also
verified at the finite chemical potential.\footnote{The critical
condition of fixed energy density  can  be  related with the
percolation condition for deconfinement \cite{satzp}.} Fig. (6)
shows recent lattice results on the position of the phase boundary
line in 2--flavor QCD obtained within the Taylor approximation and
with the pion mass $m_\pi\simeq 770$ MeV. The line of fixed energy
density $\epsilon\simeq 0.6$ GeV/$fm^3$ with $\epsilon$ calculated
in the HRG--model with the modified mass spectrum  is seen to
coincide with lattice results. Decreasing the pion mass to its
physical value and including a complete set of resonances expected
in (2+1)--flavor QCD results in the shift of the position of the
phase boundary line towards a phenomenological freeze-out
condition of the fixed energy/particle$\simeq 1$GeV.
 The splitting of freeze-out
and phase boundary line appears when the ratio of meson/baryon
multiplicities reaches the unity.

The dashed--dotted line in Fig.  (6) could  be a physical boundary
line separating the hadronic from quark--gluon plasma phase. However,
in the baryon dominating medium one expects a very efficient meson
baryon scatterings \cite{bengt}. Consequently, the particle
dispersion relations and spectral functions can be modified in a
medium.  Thus, it is not clear if in this case the partition
function of hadron resonance gas with free particle dispersion
relations is still an adequate approximation of the QCD thermodynamics
just before deconfinement. In view of the above, the position of
the critical curve shown in Fig.  (6) could be shifted in the
parameter range where $\mu_B>0.5$ GeV, that is where baryons start
to be dominating degrees of freedom in a thermal fireball.


\section{Conclusions}
We have shown that the statistical operator of hadron resonance
gas used to  describe particle yields in heavy ion collisions
provides a satisfactory description of the recent lattice results
on QCD thermodynamics at the finite chemical potential. In
particular, the basic property  of this operator like e.g. a
factorization of temperature and chemical potential dependence  is
obviously confirmed by the lattice results.   The ratios of the
coefficients in the Taylor expansion of a thermodynamic pressure
obtained in LGT are as well  described by  the expansion of
$\cosh$--function predicted by the hadron resonance gas. These
results should be independent from the particular choice  of the
quark mass in lattice calculations and are also to a large extent
free from lattice artifacts.   The phenomenological partition
function, with $m_\pi$--dependent  hadronic mass spectrum,  was
also shown to describe quantitatively the temperature and chemical
potential dependence of the basic thermodynamical observables
obtained in LGT for $T<T_c$. This indicates that the  hadron
resonance gas partition function is a good approximation of QCD
thermodynamics in the hadronic phase. Applying the above partition
function together with the condition of a fixed energy density we
have discussed a possible  position of the QCD critical curve in
the temperature--chemical potential plane.

\section*{ Acknowledgments}
Stimulating  discussions with P. Braun--Munzinger, S. Ejiri, F.
Karsch, H. Satz, J. Stachel  and A. Tawfik are kindly
acknowledged. This work was partly supported by the KBN under
grant 2P03 (06925) and by the DFG under grant GRK 881.


   \end{document}